\documentclass[3p,twocolumn]{elsarticle}

\usepackage{lineno,hyperref}
\usepackage{mathtools, cuted}
\modulolinenumbers[5]

\newcommand{\lsim}{{\;\raise0.3ex\hbox{$<$\kern-0.75em\raise-1.1ex\hbox{$\sim$}}\;}}
\newcommand{\gsim}{{\;\raise0.3ex\hbox{$>$\kern-0.75em\raise-1.1ex\hbox{$\sim$}}\;}}

\journal{Physics of the Dark Universe}

\begin{document}

\begin{frontmatter}

\title{Extended Metastable Dark Energy}


\author[IAG]{J. A. S. Lima\corref{mycorrespondingauthor}}
\cortext[mycorrespondingauthor]{Corresponding author}
\ead{jas.lima@iag.usp.br}

\author[UFABC]{G. J. M. Zilioti}
\ead{george.zilioti@ufabc.edu.br}

\author[UFLA]{L. C. T. Brito}
\ead{lcbrito@ufla.br}

\address[IAG]{Departamento de Astronomia, Universidade de S\~ao Paulo, Rua do Mat\~ao 1226, 05508-900, SP, Brazil}
\address[UFABC]{Universidade Federal do ABC (UFABC), Santo Andr\'e, 09210-580, S\~ao Paulo, Brasil}
\address[UFLA]{Universidade Federal de Lavras - Departamento de F\'{\i}sica \\
	Caixa Postal 3037, 37200-000, Lavras, Minas Gerais, Brazil}

\begin{abstract}
The  metastable dark energy scenario is revisited by assuming that the current  false vacuum energy density is the remnant from a primeval inflationary stage.  The zero temperature scalar field potential is here described by an even power series up to order six which depends on 3 free parameters: the mass of the scalar field ($m$), the dimensionless ($\lambda$) specifying the standard self-interaction term, and a free cutoff mass scale ($M$) quantifying all possible deviations from the degenerate false vacuum state. The current $\Lambda$CDM model is a consequence of the very long decay time of the false vacuum which although finite is much greater than the current age of the Universe. This result remains valid for arbitrary combinations of the $m/M$ ratio which can analytically be determined in the thin-wall approximation and numerically calculated outside this limit. Unlike many claims in the literature the vacuum dominance may be temporary. The finiteness of the decay time suggests that the ultimate stage of the observed Universe in such a scenario will not be driven by a de Sitter type cosmology.

\end{abstract}
\begin{keyword}
\texttt{Dark energy}\sep \texttt{Vacuum decay}\sep \texttt{Cosmological constant}
\end{keyword}

\end{frontmatter}

\section{Introduction}

Since the discovery of cosmic acceleration about two decades ago based on Supernovae type Ia observations,  there are plenty of efforts in the astro-particle-physicist community to determine its cause. Dark energy, as it is called the agent responsible for the unexpected late time accelerating process, has already many candidates proposed in the literature \cite{PR03,Lima,Hut}. The most successful one so far is the effective cosmological constant ($\Lambda$) defining an effective constant vacuum energy density ($\rho_V = \Lambda/8\pi G$). Its current value,  $\rho_V \sim 10^{-47}GeV^{4}$, is in agreement with a plethora of  high quality cosmological data (SNe Ia, cosmic background radiation (CMB), galaxy clustering, baryon acoustic oscillations (BAO), weak gravitational lensing, etc.). 

It is now commonly accepted that the constant $\Lambda$-term provides the most prominent and exotic piece of information underlying the current cosmic concordance model ($\Lambda$CDM). However, the interest on different forms of dark energy (quintessence, K-essence, X-matter, decaying $\Lambda$, etc), some physical mechanism, emulating the $\Lambda$CDM dynamics at the background and perturbative levels \cite{Hut,LJF2010,Waga2014, LB2014}, or even alternative gravity theories, have not declined yet (for a review see \cite{ST2019}). 
On general grounds, apart the so-called small scale problems \cite{Bullock2017,Del2014}, the  $\Lambda$CDM model have difficulties from two distinct origins. The current vacuum energy density is plagued by two cosmological puzzles, namely: the cosmological constant (CC) and the coincidence problems \cite{Weinberg:1988cp, Zlatev:1998tr, GSL2018}. 

Nevertheless, since the current $\Lambda$CDM model is well accepted by cosmologists and astronomers as the best description of the  present day Universe, it should be somehow better justified even whether such problems are not completely solved. There are some  fundamental attempts involving extra dimensions based on supersymmetry (SUSY) and also in the landscape string theory. The former is not an exact symmetry of nature and gave origin to several supergravity models based on the idea that the 4-dimensional vacuum may curve the extra dimensions \cite{Burguess}. The later is also an interesting possibility\footnote{The so-called string gas cosmology has also several interesting and testable consequences to the primeval  Universe, some of them fully distinguishable from many models of early inflation \cite{Brandenberger2015}.} to solve the CC problem. However, moduli fields are not observed and its basic solution may have about $e^{500}$ vacuum states. Hence there are doubts whether the theory is actually falsiable \cite{DouglasKachru2007,Copeland2016}. Seemingly, the problem is the nonexistence of a mechanism or selection criteria linking the SUSY breaking scale or the string landscape possibilities with the  cosmic vacuum scale now observed.   

In the  phenomenological front, the simplest possibility is that the current accelerating Universe is driven by a long lived (quintessence) false vacuum state, a possible remnant of a primeval inflationary stage \cite{Ratra,CALS2006}. In this concern, an interesting scenario dubbed metastable dark energy (MDE)  was  recently proposed  by Landim and Abdalla (from now on referred to as LA paper \cite{Landim:2016isc}). The model is based on quantum tunneling from a false to the stable true vacuum state. The decay process of the metastable state occurs in the current low energy Universe instead of at early times, as happens in the old inflationary scenario originally suggested by Guth \cite{G1981,G1983}. 

This kind of late time decaying vacuum process has at least two interesting features: (i) the low temperature of the vacuum-matter phase implies that it can be discussed  based on the semiclassical approach  developed long ago by Coleman and collaborators \cite{Coleman:1977py, Callan:1977pt}, and (ii) since none reheating mechanism is required to operate in the present phase of the Universe, this means that the current low energy inflation is not plagued by any kind of ``graceful exit" problem \cite{RB2010}.   

The dark energy model proposed  here  is powered by a scalar field whose potential is described by a power series of even self-interacting contributions up to order six  which depends on 3 free parameters: the mass of the scalar field ($m$), the numerical value of ($\lambda$) modulating the standard $\phi^{4}$ term, and a free cutoff mass scale ($M$) quantifying all possible deviations from the degenerate false-vacuum states. As we shall see, this extended model has different predictions of the LA paper. In particular,  when the false-vacuum energy density is pin down by the current observations and a value of $\lambda$ is given, only the  $m/M$ ratio can analytically be calculated in the thin-wall approximation. In addition, since the  decay rate per unit volume also depends on the ratio $m/M$, a lower limit  to the mass $m$ cannot be determined even in the thin-wall domain.

It should be remarked that for $M=M_{Planck}$ the potential adopted in \cite{Landim:2016isc} is recovered. However, some results are conceptually different. Actually, for a given value of $\lambda$, the mass of the scalar field in the thin-wall approximation becomes exactly determined under the proviso that the  thickness  of the wall is equal to the false vacuum barrier. This result suggests that the corresponding  lower limit for $m$, inferred by comparing the false vacuum decay time with the age of the Universe, $T_{DV} \geq H_0^{-1}$, is somewhat meaningless in the thin wall limit. We also give one step further by analysing  the results within the thick wall limit.

 This article is organized as follows. In section \ref{model}, we present the basics of the model, as an extension of the LA scenario. In section \ref{TWA},  a rigorous calculation of the barrier thickness,  as a function of ratio $m/M$ is accomplished assuming the thin-wall approximation. In section \ref{VDR}, the decay rate per unit volume is  calculated and compared to the present age of the Universe.  In section \ref{thick}, we perform the needed numerical calculations in order to extend the analysis beyond the thin wall limit. An accurate  fitting formula describing the thick wall results is also proposed. Finally, the article is closed in section \ref{conclusion} with a summary of our main conclusions. 

\section{The Extended MDE Model}\label{model}

To begin with, let us consider a scalar field $\phi$ with a Lagrangian density given by:
\begin{equation}
\mathcal{L}=\frac{1}{2}(\partial_{\mu}\phi)^{2} - V(\phi),  
\end{equation}
\noindent where the potential $V(\phi)$ contains a sum of even self-interactions up to order six:
\begin{equation}\label{VScalar}
V(\phi)=\mathcal{V}_0 + \frac{m^2}{2}\phi^2\left[1 - \frac{\lambda}{4m^2}\phi^2\right]^2 -\frac{1}{M^2}\phi^6.
\end{equation} 
The constant $\mathcal{V}_0$ is the value of $V(\phi =0)$, the arbitrary origin of the potential, $m$ is the mass of the field, whereas $\lambda$ is a positive dimensionless free parameter of the theory whose value is of the order $\lambda \sim 0.1$ or less. The mass $M$  is an arbitrary cutoff, for the moment satisfying only the inequality, $M\leq M_{Planck}$, while the lower limit it will be determined next as a physical constraint.  As in reference \cite{Landim:2016isc}, note that the coefficient of the first $\phi^{6}$ term was chosen so that the sum of the second, third and fourth terms becomes a perfect square.

As it appears, the $\phi^{6}$ correction is the simplest high order interaction for a non renormalizable scalar field potential. It should be interpreted in the context of effective theories \cite{Weinberg78,Weinberg08}. This non renormalizable interaction comes from  high energy field theory, its effect being regulated by the mass $M$. Several applications of corrected inflationary potentials through effective field theory are well know in the literature (see, for instance, \cite{Weinberg08, Martin13}). In \cite{Bodeker2005} it was used as the Higgs potential in a baryogenesis model. This kind of potential has also been used in many works in condensed matter. A $\phi^6$ correction was adopted by \cite{Bergner2003} to analyse the bubble dynamics formation after nucleation (see also \cite{Widyan2008} and references therein for others examples).

In what follows, it proves convenient to rewrite the above potential in terms of dimensionless quantities:
\begin{equation}\label{VScalar1}
U(\psi)=\mathcal{U}_0 + \frac{1}{2}\psi^2\left[1 - \frac{\lambda}{4}\psi^2\right]^2 - \sigma \psi^6,
\end{equation} 
where $\psi=\phi/m$, $\sigma =  m^{2}/M^{2}$, $U(\psi) = V(\phi)/m^{4}$ and $\mathcal{U}_0 = \mathcal{V}_0/m^4$. The quantity $\sigma$ works like a correction  to the normalized symmetric case.  When $\sigma = 0$, there are three degenerate minima, and, as such, the quantum tunneling process is fully suppressed. 

As one may check, the values of $\psi$ defining the extreme points of the dimensionless potential (\ref{VScalar1}) are: 
\begin{equation}
\label{minimum0}
\psi_0=0,
\end{equation}

\begin{equation}\label{minimum}
\psi_{\pm}^2 = \frac{8}{3\lambda(1-\sigma/\sigma_m)}\left[1 + \frac{1}{2}\sqrt{\left(1+3\frac{\sigma}{\sigma_m}\right)}\,\right],\\
\end{equation}

\begin{equation}\label{maximum}
\psi_{a,b}^2 = \frac{8}{3\lambda(1-\sigma/\sigma_m)}\left[1 - \frac{1}{2}\sqrt{\left(1+3\frac{\sigma}{\sigma_m}\right)}\,\right],\\
\end{equation}

where $\psi_{0}$ and $\psi_{\pm}$ are three minimal points while $\psi_{a,b}$ are two maximum points. The quantity $\sigma_m \equiv {\lambda^2}/{32}$ is the maximum value $\sigma$ can take, above that $\psi_{\pm}$ is complex valued and no longer a minimum. Real values are obtained for  ${\sigma}/{\sigma_m} \leq 1$ which can be translated as  a lower bound  on  the value of  $M$. Hence, by assuming the ``safe extreme upper bound"  the cutoff mass is constrained upon the interval: 
\begin{equation}\label{Mlimits}
4\sqrt{2}\,\,m/\lambda \leq M\leq M_{Planck}. 
\end{equation}
 It is also interesting to define the normalized dimensionless quantity representing the energy density difference between the false and true stable vacuum states\footnote{Since the scalar field is symmetrical, the system can tunnel to either one of the minima ($\psi_{\pm}$).}
 
\begin{equation}\label{eq:epsilon}
\Delta U \equiv U(\psi_0) - U(\psi_{\pm}) = \mathcal{U}_0 - U(\psi_{\pm}).
\end{equation}
Note also that $\mathcal{U}_0$, the value of the potential to the central minimum  ($\psi_0=0$), is fully independent  of $\sigma$. In other words, the $\sigma$-corrections  modify only the values of the symmetrical minima,  $U(\psi_{\pm})$. This means  that $\Delta U$ varies only because of $U(\psi_{\pm})$. For arbitrary values of the ratio $\sigma/\sigma_m$ it follows that

\begin{equation}\label{epsilon}
\Delta U = \frac{4\left[2+\sqrt{\left(1+\frac{3\sigma}{\sigma_m}\right)}\right]\left[\frac{3\sigma}{\sigma_m}-1+\sqrt{\left(1+\frac{3\sigma}{\sigma_m}\right)}\right]}{27\lambda \left(1-\frac{\sigma}{\sigma_m}\right)^2}.
\end{equation} 
 Note that for $\sigma=0$ one finds $\Delta U=0$, but,  in the limit $\sigma \rightarrow \sigma_m$ it diverges ($\Delta U \rightarrow \infty$). In our scenario the energy density difference between the false and true vacuum states may be  extremely large. Due to the dependence on the ratio $m/M$,  the thin-wall approximation must be carefully discussed.

In \textbf{Figure 1}, we display the dimensionless potential $U(\psi)$ for some selected values of $\sigma$. The case $\sigma=0$ (solid black line) is the degenerate potential as described by equations (\ref{VScalar})-(\ref{maximum}).  The dotted lines show the potential behavior for some nonzero values of the ratio $\sigma/\sigma_m$. 

\begin{figure}[ht]
	\includegraphics[width=\linewidth]{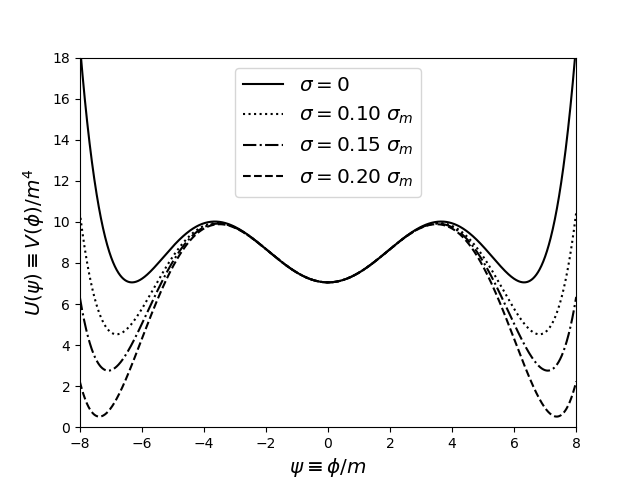}
	\caption{The zero temperature potential for a first order phase transition through quantum tunneling as defined by (\ref{VScalar}) and (\ref{VScalar1}). The solid curves describes the 3-degenerate false vacuum states  with $\lambda = 0.1$ and $\sigma=0$. The dotted lines display  the results including the correction described by $\sigma=m^2/M^{2}\neq 0$.} 
	\label{fig:potential}
\end{figure}

\section{Thin-Wall Approximation}\label{TWA}

Let us now  consider the zero temperature scalar field  trapped in the false vacuum at $\psi_0 = 0$ with a nonnull tunnelling probability (\textbf{Figure 1}). From now on, without loss of generality, the minimum $\psi_{+}$ is  chosen to be the true stable vacuum state after the tunneling process. Following standard lines, the decay rate per unit volume reads\footnote{Coleman used the notation $\Gamma/V$ \cite{Coleman:1977py}.}:

\begin{equation}\label{DR1}
\Gamma = A e^{-B},              
\end{equation}          
where  the value of $A$  with  dimension $[mass]^{4}$ is calculated from Gaussian functional integrals around the instanton solution. However, for a while we are not interested  in its exact form. Due to the exponential factor, $A$ is largely subdominant in the thin-wall approximation.  Thus, we only consider that $A \simeq m^4$, the natural energy scale of the  problem. The quantity $B = S_E$, is the Euclidean action of the instanton (the action with imaginary time, $t \rightarrow -i\tau$) 
\begin{equation}
S_E = S_E(\phi_b), 
\end{equation}
where $\phi_b$ is the solution of the Euclidean equations of motion with appropriate boundary conditions. At zero temperature, the scalar field evolution is dominated by quantum tunneling process which has $O(4)$ symmetry in the coordinate  $\rho^2 = x^2+y^2+z^2+\tau^2$. The Euclidean action can be rewritten as \cite{Coleman:1977py, CGM1978,KT1990}
\begin{eqnarray}\label{action}
S_E = 2\pi^2 \int_0^{\infty} d\rho \rho^3 \left[\frac{1}{2}(\partial_{\rho}\phi)^2+V(\phi)\right],
\end{eqnarray}
with the equation of motion taking the form:

\begin{equation}\label{eom}
\partial_{\rho}^2 \phi+ \frac{3}{\rho}\partial_{\rho}\phi - V'(\phi) = 0,
\end{equation}
which must be solved for $\phi=\phi(\rho)$ with boundary conditions $\phi(\infty) = \phi_+$, $\phi(0) = \phi_0$, $\partial_{\rho}\phi(0) = 0$.

In general, for a given potential, the bounce solution is numerically calculated. However, in the so-called  thin wall approximation, it is possible to  obtain a closed form for $S_E$. The idea is that the difference between the true and false vacuum, $\Delta V>0$ is small: 

\begin{equation}
V(\phi) \approx \tilde{V} (\phi) + \epsilon,
\end{equation}
where $\epsilon$, the small energy density difference, is the thickness of the wall, and $\tilde{V}(\phi)$ is the potential  with degenerate minima. In our case this happens for small values of the ratio $\sigma/\sigma_m$ (see \textbf{Figure 1}). As one may check from (\ref{epsilon}),  in the limit $\sigma \ll \sigma_m$, the first order correction defining  $\epsilon$ reads:  
 \begin{equation}\label{TW}
 \epsilon = \Delta V = m^{4}\Delta U \simeq \frac{2m^4}{\lambda}\frac{\sigma}{\sigma_m} = \frac{64m^6}{\lambda^3 M^2}.
 \end{equation}

Before discussing the physical meaning of the above result, let us calculate the bubble radius $a$ in order to obtain the analytical expression of $S_E$ in the thin wall limit. The integration of (\ref{action}) yields a function of the radius:
\begin{equation}\label{action1}
S_E =  -\frac{\pi^2}{2}\,a^{4}\epsilon + {2\pi^{2}}\,m^{3}a^{3} \bar{S}_1,
\end{equation}
where the ``normalized surface tension" $\bar{S}_1$ is here defined by:
\begin{equation}\label{NT}
\bar{S}_1=\int_{\psi_0}^{\psi_+} d\psi\,\sqrt{2[{\bar U}(\psi_0) - {\bar U}(\psi_+)]} = \frac{1}{\lambda}.
\end{equation}
Now, by inserting ${\bar S}_1$ into (\ref{action1}) and minimizing the result ($dS_E/da = 0$), the radius of the bubble reads: 
\begin{equation}\label{BR}
a =  \frac{3m^3}{\lambda \epsilon}.
\end{equation}
It thus follows that $a\, \propto\, \epsilon^{-1}$ so that big bubbles are associated with smaller wall thicknesses. On the other hand,  since the relevant  dimensionless quantity is $ma$,  a more precise condition for the validity of the thin-wall approximation may be defined:

\begin{table}[hbt!]
	\caption{Basic quantities from equation (\ref{TW}) for $\lambda=0.1$ and $\epsilon =10^{-47}GeV^4$. The first line are the results for $M=M_{Planck}$, the extreme limit within the thin-wall approximation. Note that for smaller values of $M$ the thin-wall is still valid  by many orders of magnitude ($\sigma \ll \sigma_m$). However, for the boldface line the results are only approximated because the thick-wall regime need to be considered (see text).}
	\centering
	
\begin{tabular}{lll}
		
		\hline 
		
		$M$ $(GeV)$ & $m$ $(GeV)$ & $\sigma/\sigma_m$ \\ 
		
		\hline
		\hline
		
		$10^{19}$  & $ 5.00 \times 10^{-3}$ & $8.00 \times 10^{-40}$  \\
		$10^{8}$  & $ 1.08 \times 10^{-6}$ & $3.71 \times 10^{-25}$  \\
		$10^{2}$  & $1.08 \times 10^{-8}$ & $3.71 \times 10^{-17}$  \\
		$1$  & $2.32 \times 10^{-9}$ & $1.72 \times 10^{-14}$  \\
		$10^{-5}$  & $5.00 \times 10^{-11}$ & $8.00 \times 10^{-8}$  \\
		$\pmb{10^{-10}}$  & $\pmb{1.07 \times 10^{-12}}$ & $\pmb{3.71\times 10^{-1}}$  \\
		\hline
	\end{tabular}
	\label{tab}
\end{table}

\begin{equation}
ma \gg 1 \Longleftrightarrow \frac{\lambda \epsilon} {3 m^4} \ll 1,
\end{equation}
and from equation (\ref{TW}), we see that the thin-wall approximation is valid only for $\sigma/\sigma_m \ll\, 1$. 

At this point, it is interesting to comment on the main consequence of the expression (\ref{TW}) defining  $\epsilon$. It implies that the mass of the scalar field can be expressed as:

\begin{equation}\label{eq:mass}
{m} = \frac{\sqrt{\lambda}}{2}{M^{1/3}\epsilon^{1/6}}.
\end{equation}

Now, by taking  $\epsilon = 10^{-47}\, GeV^4$ and $M = M_{Planck}$ as in \cite{Landim:2016isc}, one may check from (\ref{eq:mass}) that the mass of the scalar field reads:
\begin{equation}
m = \frac{\sqrt{10^{3}\lambda }}{2} \, MeV.    
\end{equation} 
Hence, for $\lambda = 0.1$ we find $m=5\,MeV$ while for $\lambda=10^{-3}$ we get $m=0.5\,MeV$. The important lesson here is that for $M = M_{Planck}$ the mass of the scalar field depends only on the $\lambda$ parameter.

Nevertheless,  since  $M$ is now a free parameter of the theory, there is no apriori reason to pin down it as the Planck mass.  Actually, we see from Eqs.  (\ref{Mlimits}), (\ref{TW}) and (\ref{BR}) that $M=M_{Planck}$ is only an extreme value for $M$ associated to the thinnest possible wall of the bubble materialized at the end of the tunneling process, that one having the biggest radius (see next section). In principle, such a freedom  may clarify  some conceptual aspects of the metastable dark energy model. In particular, as we have seen, it permits a more rigorous definition of the thin-wall approximation in this context. 

In \textbf{Table 1}, by choosing $\lambda = 0.1$ we display some values of the pair $(M,\,m)$ and also the corresponding ratios $\sigma/\sigma_m$. In the first line, apart the value for the mass $m$,  we see the results for the value of $M$ chosen in \cite{Landim:2016isc}. Note that all values of $M$ are in agreement with the constraint (\ref{Mlimits}). However, for $M=10^{-10}GeV$  we are clearly out of the limits defining the thin-wall approximation because the ratio $\sigma/\sigma_m \sim 0.37$. Of course, analytical predictions  based on these values (boldface line) cannot be taken seriously because the system is already within the thick-wall regime and, as such, the calculations must be numerically performed. 

Let us now  close this section combining the previous results in order to obtain the stationary action $S_E$ in the absence of gravity. By inserting Eqs. (\ref{TW}), (\ref{NT}) and (\ref{BR}) into (\ref{action1}) we find:

\begin{equation}\label{TWF}
S_E = \frac{27\pi^2 m^{12}}{2\lambda^4 \epsilon^3} = \frac{27\pi^2}{16\lambda} \left(\frac{\sigma}{\sigma_m}\right)^{-3}.
\end{equation}

\section{Decaying Vacuum and Hubble Time}\label{VDR}

The main aim of this section is to establish under which conditions  the
characteristic decaying vacuum time ($T_{DV}$) for the metastable false vacuum state is greater or at least equal to the current age of the Universe ($T_U$). From the above least-action (\ref{TWF}) and $A \sim m^{4}$, we rewrite (\ref{DR1}) as

\begin{equation}
\Gamma = m^{4}\,e^{-\frac{27\pi^2}{16\lambda} \left(\frac{\sigma}{\sigma_m}\right)^{-3} }\,.   
\end{equation}

It should be remarked that the constraint $T_{DV}/T_U \geq 1$ was adopted in \cite{Landim:2016isc} for obtaining a lower bound on the value of $m$ [see equation (19) there]. They found   

\begin{equation}\label{mL}
m \geq 10^{-12} \; GeV. 
\end{equation}
Further, by using the above mass constraint and Eq. (\ref{BR}), a lower bound  to the  bubble radius, $a \geq 0.03\; cm$, was also obtained.

There is, however, a doubtful aspect concerning the validity of the quoted lower bound for $m$.  As in the previous section, the choices $\epsilon = 10^{-47}\, GeV^4$ and $M = M_{Planck}$ yield, for $\lambda =0.1$,  a mass of the scalar field $m = 5 \; MeV$. Of course, the above lower bound is in agreement with this value of the mass. However, one may argue that it is somewhat ambiguous because the action was already completely defined. In other words, there is no more freedom to define a lower bound on the mass $m$. Actually, from equation (\ref{TWF}) we find:

\begin{equation}\label{ActionLA}
S_E = \frac{27\pi^2 m^{12}}{2\lambda^4 \epsilon^3} = \frac{27\pi^2}{16\lambda} \left(\frac{\sigma}{\sigma_m}\right)^{-3}= \frac{27\pi^2\times 10^{120}}{16\lambda}\,,
\end{equation}
\begin{table}[hbt!]
	\caption{The bubble radii and  the  decaying time of the false vacuum compared with the current age of the Universe ($\Gamma/H_{0}^4$) for the selected  $\sigma/\sigma_m$ values listed on {\bf Table 1}.} 
	\centering
	
\begin{tabular}{lll}
		
		\hline 
		
		$\sigma/\sigma_m$ & Radius (cm) & $\Gamma/H_0^4$  \\ 
		
		\hline
		\hline
		
		$8.00 \times 10^{-40}$ & $7.42 \times 10^{27}$ & $\exp(-3.25 \times 10^{119})$  \\
		$3.71 \times 10^{-25}$ & $7.42 \times 10^{16}$ & $\exp(-3.25 \times 10^{75})$ \\
		$3.71 \times 10^{-17}$ & $7.42 \times 10^{10}$ & $\exp(-3.25 \times 10^{51})$ \\
		$1.72 \times 10^{-14}$ & $7.42 \times 10^{8}$ & $\exp(-3.25 \times 10^{43})$  \\
		$8.00 \times 10^{-8}$ & $7.42 \times 10^{3}$ & $\exp(-3.25 \times 10^{23})$  \\
		$\pmb{3.71\times 10^{-1}}$ & $\pmb{7.42 \times 10^{-2}}$ & $\pmb{\exp(-3.25 \times 10^{2.96})}$   \\
		\hline
	\end{tabular}
	\label{tab2}
\end{table}
where the value of $\sigma/\sigma_m$ was taken from the first line of \textbf{Table 1}, with $\lambda=0.1$, $M = M_{Planck}$ and $m=5Mev$.   As we shall see below, the above value of the action implies the condition, $T_{DV} \gg T_U$. In this sense, even considering that the  above limit (\ref{mL}) is somewhat misleading,  one can say that the LA model satisfies such a  natural consistency check. 

On the other hand, for a mass $m=5MeV$, the bubble radius predicted by (\ref{BR}) is $a=7.42\,\times 10^{27}$\,cm, basically, the Hubble radius today.  

In what follows, we also consider $\epsilon=10^{-47}\,GeV^4$, in order to  perform the calculations without assuming that $M$ is the Planck mass. 

To begin with let us recall that the present age of the Universe in the current $\Lambda$CDM model is exactly the inverse of the Hubble parameter, $H^{-1}_0$ \cite{Lima:2007kr}. In this way, the decay rate per unit volume should be compared to  $H_0^4$\cite{TW1992}: 

\begin{equation}\label{eq:gammaH4}
\frac{\Gamma}{H_0^4} = \frac{m^4}{H_0^4} \exp \left[-\frac{27\pi^2}{16\lambda}\left(\frac{\sigma}{\sigma_m}\right)^{-3}\right] \lsim 1.
\end{equation} 
By using the above equation as a consistency check, we may see whether the decay time for a generic $M$ can be bigger than the age of the universe. 

In \textbf{Table 2} we show the calculated bubble radius $a$ and the ratio $\Gamma/H_0^{4}$ for  $\lambda=0.1 $ and the same values of $\sigma/\sigma_m$ listed on \textbf{Table 1}. Note that except for the values shown in the bottom (boldface line), for all the remaining cases the system is well within the thin-wall approximation ($\sigma \ll \sigma_m$) for which LA model is the extreme case. In the first line we have displayed  the  results of \cite{Landim:2016isc} at light of the present treatment.  In the last column we see that the  $\Gamma/H_0^{4}$ ratio is by far much less than unit  This means that  the decay time is many orders of magnitude bigger than the present age of the universe ($T_{DV} \gg T_U$), as long as the treatment is out of the thick wall approximation. 

It is also worth notice that gravity does not change appreciably  the above result derived in the thin-wall approximation. As discussed by Coleman e De Lucia \cite{CDL1980}, the basic  general relativistic (GR) effect is to change $S_E \rightarrow \tilde S$ given by 

\begin{equation}\label{LAction}
\tilde S = \frac{S_E}{{\left[{1 + (\frac{a}{2\Delta})^{2}}\right]}^{2}},
\end{equation}
where $S_E$ is the least-action (\ref{TWF}) in the absence of gravity,  $a$ is given by equation (\ref{BR}) and $\Delta  = {\sqrt{3/\epsilon}}\,\,M_{Planck}$ is the Schwarzschild radius associated to  a sphere of energy density $\epsilon$, the false vacuum energy density. In the present case, as long as the thin-wall remains valid, it is also easy to show that  GR effects are negligible. In order to see that, let us calculate the correction. As one may check, it is given by:

\begin{equation}
\delta = (\frac{a}{2\Delta})^{2}= \frac{3\lambda}{256}\,\frac{M^{2}}{M^{2}_{Planck}}.
\end{equation}
Hence, the gravity correction does not depend on $m$ and has an upper bound  $\delta \leq 3\lambda/256$. Its maximum correction occurring exactly for $M=M_{Planck}$. For $\lambda \sim 0.1$ we get  $\delta \sim 10^{-3}$. 
We should note that all the results here were obtained in the thin wall approximation so that by relaxing this assumption the results can be quite different. 

\section{Thick Wall: Numerical Solution}\label{thick}

 The thin wall approximation is important as a closed analytical solution, however, its validity domain is somewhat limited, and, as such, definitive conclusions are not possible at this stage \cite{Samuel1991}. Now, in order to explore the behavior of the model beyond this limit (the so called thick wall domain), we are forced to seek a complete numerical solution. The algorithm to solve the equation of motion from the action \eqref{action} is relatively simple (see \cite{Wainwright1991} for the numerical code and details).

\begin{figure}[ht]

	\includegraphics[width=\linewidth]{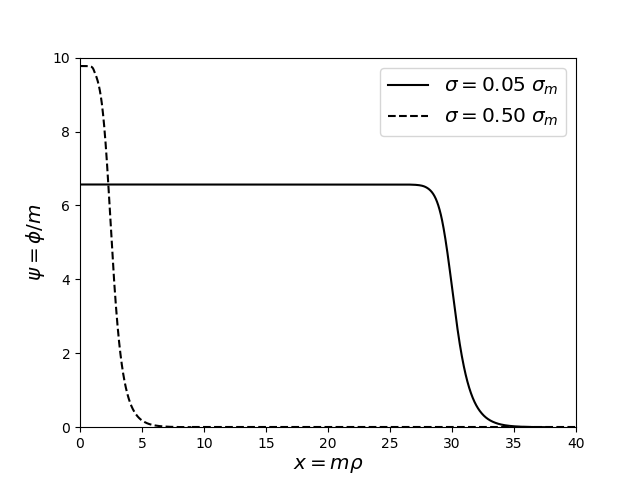}
		\caption{Profile of two possible bubbles, one for $\sigma = 0.5\sigma_m$ and other for $\sigma = 0.05\sigma_m$. Solid line is the thin wall case, while dashed is thick.}
		\label{fig:perfilbolha}
\end{figure}

In {\bf Figure \ref{fig:perfilbolha}}, we show the profile of two bubbles as numerically obtained. The solid line is the result for the thin wall approximation, while the dashed line is a thick wall bubble. Note that the characteristic radius of the bubble is considerably larger than its wall in the thin case, while the corresponding sizes have the same order in the thick one.

to extend the analytical thin wall solution for all values of $\sigma/\sigma_m$. The idea is to consider the result \eqref{TWF} as the zeroth order expansion of a more complete equation. The fitting formula can be written as:

\begin{align}\label{fitting}
	S_E &\rightarrow \bar{S}_E = \frac{27\pi^2}{16\lambda} \left(\frac{\sigma}{\sigma_m}\right)^{-3} F\left(\frac{\sigma}{\sigma_m}\right),
\end{align}
 where the arbitrary function $F({\sigma}/{\sigma_m})$ is expanded in a Taylor series until third order
\begin{equation}\label{fitting1}
F= \left[1+p\left(\frac{\sigma}{\sigma_m}\right)+q\left(\frac{\sigma}{\sigma_m}\right)^2+r\left(\frac{\sigma}{\sigma_m}\right)^3 + \cdots \right],
\end{equation}
 with the dimensionless parameters $p$, $q$ and $r$ numerically fitted. The corresponding results are:
\begin{align}
	p &= 2.648, \nonumber \\
	q &= 2.997, \\ 
	r &= -4.503. \nonumber
\end{align}

\begin{figure}[ht]
	\includegraphics[width=\linewidth]{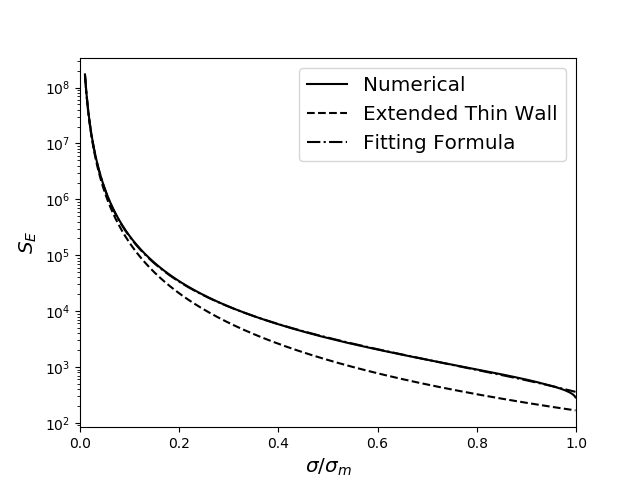}
	\caption{Comparison between the value of $S_E$ for thin wall, numerical and fitting formula. Note that the numerical and fitting formula curves are on top of each other.}
	\label{fig:numerico}	
\end{figure}

 In {\bf Figure \ref{fig:numerico}}, we compare the results of the different approaches: (i) numerical, (ii) thin wall, and (iii) fitting formula. We see the region where the thin wall regime  $\sigma \ll \sigma_m$) is valid. Note its difference for the thick wall regime and how our analytical fitting formula  describes accurately the numerical result.

 In \textbf{Tables 3} and \textbf{4}, we extend the analysis of the previous sections to the thick wall case. Instead of using equation \eqref{TW} to calculate $\sigma/\sigma_m$, we consider the complete equation \eqref{epsilon} without approximations.

\begin{table}[hbt!]
	\caption{Basic quantities from equation \eqref{epsilon} for $\lambda=0.1$ and $\epsilon =10^{-47}GeV^4$ in the thick wall regime. The first line is the same as the last line in \textbf{table 1} but using the full numerical calculation.}
	\centering
	
	\begin{tabular}{lll}
		
		\hline 
		
		$M$ $(GeV)$ & $m$ $(GeV)$ & $\sigma/\sigma_m$ \\ 
		
		\hline
		\hline
		
		$10^{-10}$  & $ 9.51 \times 10^{-13}$ & $0.289$  \\
		$10^{-10.5}$& $ 4.75 \times 10^{-13}$ & $0.715$  \\
		$10^{-11}$  & $ 1.73 \times 10^{-13}$ & $0.955$  \\
		$10^{-12}$  & $ 1.76 \times 10^{-14}$ & $0.999$  \\

		\hline
	\end{tabular}
	\label{tab3}
\end{table}  

\begin{table}[hbt!]
	\caption{The bubble radii and  the  decaying time of the false vacuum compared with the current age of the Universe ($\Gamma/H_{0}^4$) for the selected  $\sigma/\sigma_m$ values listed on {\bf Table 3}, in the thick wall regime.} 
	\centering
	
	\begin{tabular}{lll}
		
		\hline 
		
		$\sigma/\sigma_m$ & Radius (cm) & $\Gamma/H_0^4$  \\ 
		
		\hline
		\hline
		
		$0.289$ & $0.1055$ & $\exp(-12922)$  \\
		$0.715$ & $0.0586$ & $\exp(-996.24)$  \\
		$0.955$ & $0.0499$ & $\exp(-179.69)$ \\
		$0.999$ & $0.0899$ & $\exp(-19.63)$ \\

		\hline
	\end{tabular}
	\label{tab4}
\end{table}

 The first lines in \textbf{Tables 3} and \textbf{4} should be compared to the bold lines in \textbf{Tables 1} and \textbf{2}. By performing the complete calculation, we see that the value of $\sigma/\sigma_m$ is slight lower than before. As $M$ decreases $\sigma$ quickly goes to $\sigma_m$, and, even though the decay rate is rapidly increasing, the decay times are still orders of magnitude bigger than the Hubble time. 
 
 At this point, one may ask about the future of the Universe in the framework of this extended metastable vacuum decay model. In the standard view, the transition to the new phase evolves through rapidly expanding nucleation of true vacuum bubbles inside the false vacuum, a process driving the whole Universe to the true vacuum state. If the transition of the long lived false vacuum is successful in an extremely low energy environment, the model suggests a decelerating expansion in the future, driven by cold dark matter plus baryons. We recall that some cosmographic studies based on SNe Ia suggest that cosmic acceleration could already have peaked and be presently slowing down, which would imply that the recent accelerated expansion of the universe is a transient phenomenon \cite{GL2011} (for a more general approach see \cite{JUN2012}.
 
 Nevertheless, some points still need to be considered. Firstly, it should be recalled that the true vacuum was assumed here to be a zero cosmological constant ($VEV \equiv 0$), and, as such, a de Sitter like Universe in the future is unlikely unless such a condition is relaxed. Naturally, if the $VEV \neq 0$ a new de Sitter phase still remains as a possibility. Some authors have also recently conjectured that there are solution where bubbles of true vacuum inside a Universe of false vacuum will not grow \cite{Gonzales2018}. In this case, there is no bubble expansion that would convert the false vacuum into the true one.
 
 On the other hand, in the original metastable model, the scalar dark energy field was also embedded into a dark sector extension of the standard model (SM) with SU(2)R symmetry \cite{Landim:2016isc}. It was assumed that dark energy and dark matter are doublets under SU(2)R and singlets to other symmetries. In addition, also considering that the dark sector interacts with the SM particles only through gravity, the authors concluded that the decay products are compatible with a late time cosmology endowed with dark energy-dark matter interaction, as long as the coupling in the hidden sector is proportional to the Hubble parameter. Naturally, this is an interesting possibility which also deserves a closer scrutiny. In this concern, it should also be recalled that the influence of gravity in the decaying process is safely negligible only in the thin wall approximation. As above discussed, further investigation is needed for the thick wall domain when only numerical solutions are available (see \textbf{Tables 3} and \textbf{4}).

\section{Conclusion}\label{conclusion}

In this article we have proposed  an extension of the metastable dark energy model for describing the current accelerating stage of the Universe, a quantum tunneling event activated by the materialization of a bubble of true vacuum within the false vacuum.  As we have seen, the extended model depends on 3 free parameters: two mass scales  ($m,M$) and  the dimensionless $\lambda$ associated to the $\phi^4$ self-interaction.  

The ratio $m/M$ was analytically determined in the thin wall approximation by adjusting the thickness of the barrier to the current false vacuum energy density ($\epsilon \sim 10^{-47}\,GeV^{4}$). Given such a ratio, the time decay rate (per unit volume) of the false vacuum was demonstrated to be finite and much greater than the current Hubble time (see \textbf{Table 2}). 

In the thick wall regime we performed numerical calculations for the same value of the current vacuum energy density.  It was shown that the decay rate, albeit lower, is still bigger than the Hubble time, even when we reach the limit $\sigma \rightarrow \sigma_m$ (see \textbf{Table 3}). A simple analytical fitting formula describing  the numerical solution with great accuracy was also proposed [see Eqs. ($\ref{fitting}$)-($\ref{fitting1}$)].

In principle, the ‘‘friction term’’, temperature effects, and the expansion rate are negligible in the present low energy stage of the Universe, but quantum corrections should be considered. In addition, given that a ‘‘graceful exit’’ is not required here and by assuming that such a process is realistic to some degree, it means that the end of cosmology will not be driven by a de Sitter type spacetime. Actually, after the false vacuum decaying process, the present scenario also seems to be compatible with a late time decelerating Universe driven only by nonrelativistic matter as phenomenologically suggested by some authors \cite{CALS2006,Lima:2007kr}.

Naturally, whether the $\phi^6$ potential discussed here is also applied with temperature corrections to the early inflationary scenario, all these effects should be taken into account and the result should be quite different. In the same vein, it is interesting to see if the early and late time inflationary processes are somehow related. It should be recalled that at extremely high temperature the favoured phase is symmetric and probably populated by effectively massless particles. In principle, this unsettled evolutionary mystery uniting the early and late time Universe in this context does not depend only on standard energetic considerations, and, as such, it deserves a special attention for any kind of microscopic model describing the dark sector.

\vskip 0.1cm
\noindent \textbf{Acknowledgments}
\vskip 0.1cm
This work was supported by  CNPq, CAPES (PROCAD\,2013) and FAPESP (LLAMA Project).

\end{document}